\documentclass[journal]{IEEEtran}

\ifCLASSINFOpdf

\else

\fi

\usepackage{amsmath}
\usepackage{amssymb}
\usepackage{mflogo,texnames}
\usepackage{indentfirst}
\usepackage{graphicx}
\usepackage{abstract}
\usepackage{algorithm}
\usepackage{algorithmic}
\usepackage{multirow}
\usepackage{microtype}
\usepackage{color}
\usepackage{verbatim}
\usepackage{url}
\usepackage{cite}
\begin{document}

\title{Functional connectome fingerprinting: Identifying individuals and predicting cognitive function via deep learning}

\author{Biao~Cai$^{a}$, Gemeng~Zhang$^{a}$, Aiying~Zhang$^{a}$, Li~Xiao$^{a}$, Wenxing~Hu$^{a}$, Julia~M.~Stephen$^{b}$, Tony~W.~Wilson$^{c}$\\ Vince~D.~Calhoun$^{d}$ and Yu-Ping~Wang$^{a}$\\
$^{a}$Biomedical Engineering Department, Tulane University, New Orleans, Louisiana, USA\\
$^{b}$The Mind Research Network, Albuquerque, New Mexico, USA\\
$^{c}$Department of Neurological Sciences, University of Nebraska Medical Center (UNMC), Omaha, NE, USA\\
$^{d}$Tri-institutional Center for Translational Research in Neuroimaging and Data Science (TReNDS) (Georgia State University, Georgia Institute of Technology, Emory University), Atlanta, GA 30030}


\maketitle
\begin{abstract}
\par The dynamic characteristics of functional network connectivity have been widely acknowledged and studied. Both shared and unique information has been show to be present in the connectomes. However, very little has been known about whether and how this common pattern can predict the individual variability of the brain, i.e. "brain fingerprinting", which attempts to reliably identify a particular individual from a pool of subjects. In this paper, we propose to enhance the individual uniqueness based on an autoencoder network. More specifically, we rely on the hypothesis that the common neural activities shared across individuals may lessen the individual discrimination. By reducing contributions from shared activities, inter-subject variability can be enhanced. Results show that that refined connectomes utilizing an autoencoder with sparse dictionary learning can successfully distinguish one individual from the remaining participants with reasonably high accuracy (up to $99.5\%$ for the rest-rest pair). Furthermore, high-level cognitive behavior (e.g., fluid intelligence, executive function, and language comprehension) can also be better predicted using the refined functional connectivity profiles. As expected, high-order association cortices contributed more to both individual discrimination and behavior prediction. The proposed approach provides a promising way to enhance and leverage the individualized characteristics of brain networks.
\end{abstract}


\IEEEpeerreviewmaketitle

\section{Introduction}
\par Functional magnetic resonance imaging (fMRI) allows for non-invasive interrogation of brain functions based on the blood-oxygenation-level-dependent-signal (BOLD signal) \cite{biswal1995functional, greicius2003functional,damoiseaux2006consistent}. Intriguingly, a functional connectome based on functional connectivity (FC) extracted from the fMRI time series provides a promising tool to investigate individual differences in human cognitive and behavioral performance from a network perspective. Recently, numerous studies have reported individual variability in functional connectivity. For instance, Airan et al. provided an in-depth study to evaluate the degree of influence that standard fMRI acquisition and analysis schemes have on individual subject variability \cite{airan2016factors}. This variability is assumed to be associated with both genetic and environmental factors, and thereby neural development. Meanwhile, such variability also may partially affect individual cognition and behavior \cite{baldassarre2012individual, gerraty2014transfer}.

\par More importantly, the functional connectome of the human brain constitutes individualized patterns that enables us to identify one from a pool of individuals just like a fingerprint \cite{finn2015functional}. Specifically, Finn et al. demonstrated that such connectivity profiles could be used to distinguish individuals among adult participants across rest/task modalities. In their work, they showed that the discriminative subnetworks of individuals contributed most to the prediction of fluid intelligence score \cite{finn2015functional}. Kaufmann et al. reported that the functional profile developed into a stable, individual wiring pattern during adolescence, and they demonstrated that reduced mental health induced a delay and an overall reduction of such wiring \cite{kaufmann2017delayed}. The studies mentioned above used a standard procedure to extract the functional profiles, but overlooked the influence from the group-wise contribution. Inspired by this, we refined the measures of capturing individual connectomes by increasing inter-subject variability across FC values in a population with a goal of improving the power of predicting individuality with FC fingerprints \cite{cai2019refined}. Recently, the limitation of static connectivity has been widely realized, and the concept of dynamic connectivity has emerged to emphasize the time-varying characteristics of the FC \cite{hutchison2013resting, hutchison2013dynamic, allen2014tracking, calhoun2014chronnectome, calhoun2016time, cai2017estimation, cai2018capturing, zhang2019estimating}. Incorporating the information from the time-varying FC, Liu et al. studied whether and how the dynamic properties of the chronnectome acted as a fingerprint of the brain to identify individuals \cite{liu2018chronnectome}. Mounting evidence indicates that during the dynamic FC analysis, a state (termed stable state) that resembles the static FC reoccurs more frequently than others \cite{hutchison2013dynamic, allen2014tracking}. These stable states are similar across subjects and share the basic configuration with all dynamic patterns. Hence, we hypothesize that the common neural activities, which do not have the individual-specific characteristics in the connectomes, can be represented by the static FC. They may impede the revealing of individualized characterization. To our knowledge, this factor has not yet been considered when performing the individualized pattern analysis. Thus, our motivation is that, by reducing the contribution from the common neural activity, we increase sensitivity to individual variability in FC.

\par To this end, we employ a dimensionality reduction technique to extract the basic configuration. More specifically, we project fMRI data onto its underlying subspace with a common structure. A simple and commonly used method is principal component analysis (PCA), which finds the direction of the greatest variance in the dataset and represents each data point by its coordinates along each of these directions \cite{wold1987principal}. However, PCA cannot extract nonlinear structures modeled by higher than second-order statistics. Various methods have been proposed for nonlinear dimension analysis, such as the auto-associative networks, generalized PCA and kernel PCA \cite{malthouse1998limitations, kramer1991nonlinear, karhunen1995generalizations, scholkopf1998nonlinear}. The more recently proposed autoencoders (AEs) \cite{hinton2006reducing} belong to a family of nonlinear dimensionality reduction methods using neural networks. Through multi-layer neural networks, the autoencoder and its extensions demonstrate powerful performance to learn key features from data \cite{lee2008sparse, rifai2011contractive, vincent2008extracting}. For real-world datasets, successful applications include \cite{hinton2006reducing, vincent2010stacked, le2013building}. Thus, in this work, we employ autoencoder to estimate the common neural activity from rest/task fMRI data.

\par The remainder of this paper is organized as follows. In Section 2, we first describe the dataset used in this work. Then, we introduce our proposed framework for estimating FCs step by step, followed by a series of experiments conducted. In particular, we analyze whether refined connectomes extracted by our proposed method can better distinguish each individual from a pool of participants, and predict high-level cognitive behaviors. The corresponding results are illustrated in Section 3. Some discussions and concluding remarks are given in Section 4 and 5.

\section{Materials and methods}
\subsection{Data acquisition}
\par We used the publicly available S1200 Data Release of the Human Connectome Project (HCP) \cite{van2013wu}. The S1200 release contains behavioral and 3T MR imaging data from 1206 healthy young adult participants collected from August 2012 to October 2015. 889 subjects have complete data for all four 3T MRI modalities in the HCP protocol: structural images (T1w and T2w), resting-state fMRI (rsfMRI), task fMRI (tfMRI), and high angular resolution diffusion imaging (dMRI). A written informed consent was obtained for each subject. All HCP subjects were scanned on a customized Siemens 3T "Connectome Skyra" housed at Washington University in St. Louis, using a standard 32-channel Siemens receiver head coil and a "body" transmission coil designed by Siemens specifically for the smaller space available using the special gradients of the WU-Minn and MGH-UCLA Connectome scanners. To address head motion, dynamic head position information was acquired using an optical motion tracking camera system (Moire Phase Tracker Kineticor). fMRI was acquired using a whole-brain multiband gradient-echo (GE) echoplanar (EPI) sequence with the following parameters: TR/TE = 720/33.1ms, flip angle = 90$^{\circ}$, FOV = 208 $\times$ 180mm, matrix = 104 $\times$ 90 (RO $\times$ PE), multiband factor = 8, echo spacing = 0.58ms, slice thickness = 2mm. The resulting normal voxel size was 2.0 $\times$ 2.0 $\times$ 2.0mm.

\par The resting-state runs (R1 and R2) were acquired in separate sessions on two different days. Task runs included the following: working memory (Wm), motor (Mt), language (Lg) and emotion (Em). The working memory task and motor task were acquired on the first day, while the language and emotion tasks were acquired on the second day. Note that not all participants have these 6 modalities of fMRI data. Thus, we filtered out subjects lacking one or two modalities of the fMRI scanning session. Following data selection, a cohort of 862 participants (aged 22-35 years, 409 male and 453 female) was included in our analyses. Within each session, oblique axial acquisitions alternated between phase encoding in a left-to-right (LR) direction in one run and phase encoding in a right-to-left (RL) direction in another run. Here, we included only the left-to-right encoding runs to avoid potential effects of different phase encoding directions on our findings. More details about S1200 Data Release of the HCP can be found in the reference manual \cite{wu20171200}.

\subsection{Data preprocessing}
\par Our study used the fMRI dataset from HCP with the minimal preprocessing pipeline, which included gradient distortion correction, head motion correction, image distortion correction, spatial normalization to standard Montreal Neurological Institute (MNI) and intensity normalization \cite{glasser2013minimal}. Further, we applied the standard preprocessing procedures to reduce biophysical and other noise sources in the minimally processed fMRI data. These procedures contained the removal of linear components related to the 12 motion parameters (original motion parameters and their first-order derivatives), removing linear trend and performing band-pass filtering (0.01-0.1Hz). Notably, Finn et al. has reported that the smoothing level had essentially no effect on identification accuracy \cite{finn2015functional}. Thus, we investigated the analysis based on the data without applying spatial smoothing.

\par
\begin{figure}[htbp]
    \centering
    \includegraphics[width=0.48\textwidth,height=0.3\textwidth]{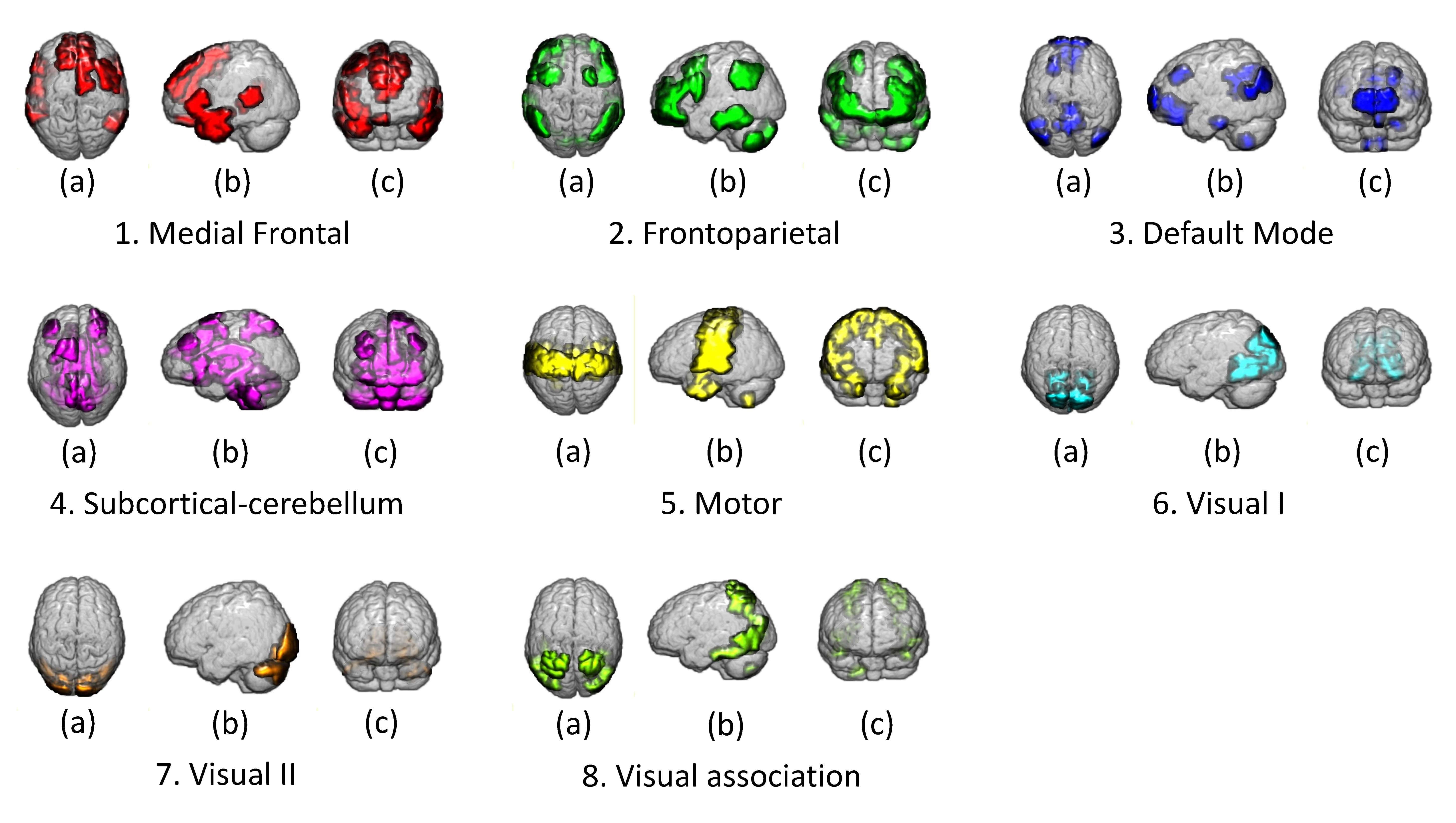}
    \caption{Axial, sagittal and coronal views (from left to right) of 8 functional networks provided by Finn et al. 1. Med F: Medial Frontal; 2. FP: Frontoparietal; 3. DMN: Default Mode; 4. Sur-Cer: Subcortical-cerebellum; 5. Mt: Motor; 6. Vis I: Visual I; 7. Vis II: Visual II; 8. Vis Assn: Visual Association.}
    \label{figure1}
\end{figure}

 To facilitate the understanding of behaviors associated with different brain regions, we applied a 268-node functional atlas provided by Finn et al.\cite{finn2015functional}, which was defined using a group-wise spectral clustering algorithm \cite{shen2013groupwise}. More specifically, we extracted the time series of each node by averaging the time courses of all voxels that belonged to that node. Then, we assigned these nodes into 8 functional networks, including medial frontal (Med F), frontoparietal (FP), default mode (DMN), subcortical-cerebellum (Sub-Cer), motor (Mt), visual I (Vis I), visual II (Vis II) and visual association (Vis Assn) regions. Axial, sagittal and coronal views of these functional networks were displayed in Fig.\ref{figure1}.

\subsection{Autoencoder network construction}
\par An autoencoder was used to extract common neural activity from the BOLD time series. For each participant of a modality, BOLD time courses with $p$ ROIs and $n_t$ time points ($p$, $n_t\in \mathbb{N}$) are available. These signals are marked as the original time series and set as inputs to the AE network. The AE network can be defined as:

\begin{equation}
    y = f_{\theta}(x) = s_1(Wx + b),
    \label{equation1}
\end{equation}
\begin{equation}
    z = g_{\theta'}(y) = s_2(W'y + b'),
    \label{equation2}
\end{equation}
where the deterministic mapping $f_{\theta}$ is called the encoder, which transforms an input vector $x\in\mathbb{R}^d$ into the hidden representation $y\in\mathbb{R}^d$. Its parameter set is $\theta = \{W, b\}$, where $W$ is a $d'\times d$ weight matrix with $d'\le d$ and $b$ is an offset vector of dimensionality $d'$.

\par
\begin{figure}[htbp]
    \centering
    \includegraphics[width=0.48\textwidth,height=0.4\textwidth]{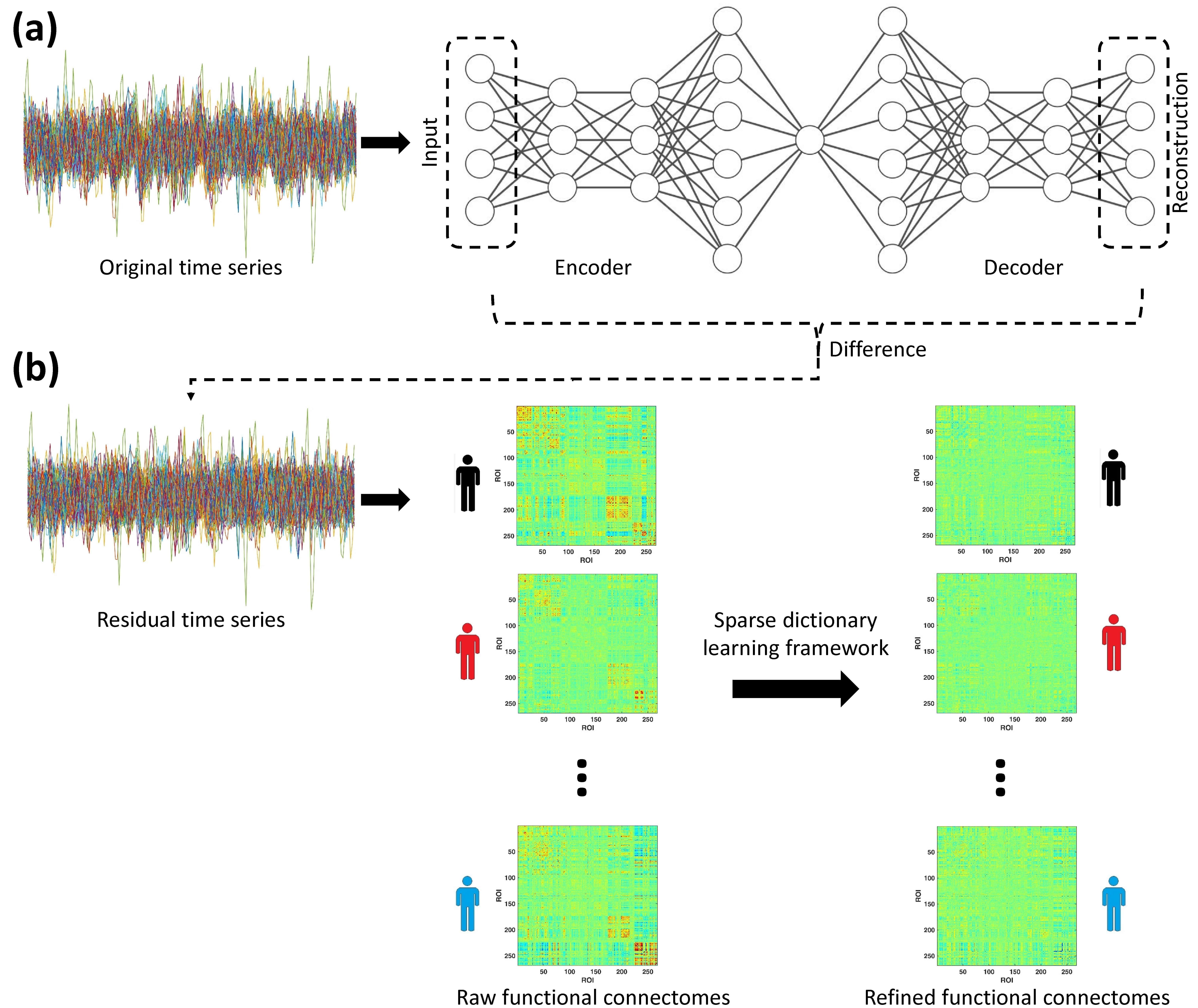}
    \caption{An illustration of the workflow to refine the brain connectivity. (a) Extraction and reduction of the effect of common neural activity using the autoencoder network. The dimensions of the AE network are set to be $d$-500-500-2000-10-2000-500-500-$d$ for each participant. The difference between the original and reconstructed time series (residual time series) is set as the input to a sparse dictionary learning model (SDL). (b) Decomposition of the FC into both the group-wise and subject-wise patterns using the SDL model. Note the assumption that the subject-specific FC may carry most of the identification information for a participant, which is tested and implemented here.}
    \label{figure2}
\end{figure}

 The resulting hidden representation $y$ is then mapped back to a reconstructed $d$-dimensional vector $z$ in the input space,i.e., $z=g_{\theta'}(y)$. This mapping $g_{\theta'}$ is called the decoder. $s_1$ and $s_2$ are activation functions for encoding and decoding layers respectively. Here, we used rectified linear units (ReLUs) in all encoder/decoder pairs, except for $s_2$ of the first pair (because the time series have both the positive and negative values) \cite{nair2010rectified}. Training was performed by minimizing the least-square $\|x-z\|_2^2$. After training of one layer, we used its output $f_{\theta}(x)$ as the input to train the next layer. In order to avoid dataset specific tuning as much as possible, we set the parameters in the DNNs (deep neural networks) training as the default (following the recommendation in \cite{van2009learning}). More importantly, inspired by the work provided by van der Maaten \cite{van2009learning}, we set layers' dimensions in AE as $d$-500-500-2000-10-2000-500-500-$d$ for all fMRI modalities, where $d$ is the number of time points (i.e., $d = n_t$) varying with respect to different fMRI modalities. All layers are fully connected. Then, we calculated the residual time courses, which are the differences between the original time series and reconstructed ones generated by the AE network. Next, the residual time series were set as the inputs to the subsequent sparse dictionary learning (SDL) model. An illustration of the workflow is displayed in Fig.\ref{figure2}(a).

\subsection{Increased individual identifiability employing the sparse dictionary learning model}
\par In our previous study, we indicated that individual connectivity analysis benefits from group-wise inferences and the refined connectomes are indeed desirable for brain mapping \cite{cai2019refined}. Thus, to further improve the inter-subject variability across FCs, we implemented the same pipeline to reduce group-wise contribution. Assume that we have $n \in \mathbb{N}$ subjects. For the residual time courses ($p$ ROIs and $n_t$ time points, $p, n_t \in \mathbb{N}$), we first calculate a correlation matrix, $C_i \in \mathbb{R}^{p\times p}(i \in 1,2,\cdots,n)$, for each subject. $C_i(b_1, b_2)$ is the Pearson correlation between ROIs $b_1$ and $b_2$ across the entire residual time series. In consideration of the symmetry of the correlation matrix, we discard the upper triangular part of $C_i$. This leads to the edge weight vector $e_i = vec(C_i) \in \mathbb{R}^{p(p-1)/2}$ for each subject. Next, we concatenate edge weight vectors from all subjects to form all the subject matrices $Y = [e_1, e_2, \ldots, e_n]$ with the size of $m \times n$, where $m = p(p-1)/2, n=1,2, \ldots, N$. Identifying the sparse representation of the functional connectivity across subjects ($Y$) can be modeled as an SDL problem. By solving the following formulation, we can approximate the given data $Y$:
\begin{equation}
\begin{aligned}
    &\min_{D,X} \|Y-DX\|_F^2 \\
    &subject\ to\ \|x_i\|_0\leq L, i = 1,2,\ldots,n,
    \label{equation3}
\end{aligned}
\end{equation}
where $L$ is a non-negative model parameter to control the sparsity level of representations. $D \in \mathbb{R}^{m\times K}$ denotes the dictionaries, and $K$ is the size of dictionaries. $X = [x_1, x_2, \ldots, x_n] \in \mathbb{R}^{K\times n}$ is the representation matrix and $\|\cdot\|_0$, $\|\cdot\|_F$ denote the $l_0$ and Frobenius norms, respectively. More details about the SDL model can be found in our previous work \cite{cai2019refined}.

\par Since we want to improve the inter-subject variability, group-wise contributions can be excluded from each correlation matrix $C_i$ to obtain a new refined functional connectome $\widehat{C}_i$. The refined functional pattern is defined as follows:

\begin{equation}
\begin{aligned}
    \widehat{C}_i = C_i-mat(Dx_i)
    \label{equation4}
\end{aligned}
\end{equation}
where $mat(Dx_i) \in \mathbb{R}^{p\times p}$ is the correlation matrix reconstructed from the lower triangular information $Dx_i$. The framework for the SDL model is illustrated in Fig.\ref{figure2}(b). Note that during analysis, the SDL model is performed on each subject of ROI network from different fMRI modalities individually.

\subsection{Individual identifiability analysis}
\par To explore the use of functional connectomes as fingerprints using our pipeline, we investigated individual identification ability proposed by Finn et al. \cite{finn2015functional}. Identification is performed across pairs of scans consisting of one target and one session from the HCP database, with the requirement that the target and database sessions are from different days to avoid the interference as much as possible. For the target session with a given subject (e.g., resting-state1, R1), we would like to identify that the connectivity pattern from the session in the database (e.g., language, Lg) belongs to the same subject. More specifically, for each participant, we first compare the correlation matrix of this subject from session 1 to each of the matrices of all the participants from session 2 ($s1\rightarrow s2$). For each comparison, the similarity scores between the connectivity patterns from session 1 and session 2 are simply estimated using the Pearson correlation coefficient. Then, we assign this participant the same label with the subject in session 2 who has the maximal similarity score with this participant. If the FCs with same label
are indeed from the same participant, the identification accuracy is considered to be $100\%$. Otherwise, it is designated as $0\%$. By calculating the proportion of subjects with the correct identification, we determine the identification accuracy of all the participants. Finally, the session 1 and session 2 are reversed, and the procedures discussed above are repeated ($s2\rightarrow s1$). Because we have three fMRI modalities (R1, Wm, Mt) for one day and three modalities of fMRI (R2, Lg, Em) for another day, this results in 9 possible combinations for $s1\rightarrow s2$ (likewise, 9 possible combinations for $s2\rightarrow s1$).

\par After obtaining the identification accuracy for all the participants, we performed 10,000 nonparametric permutation tests (two-sided) to assess whether the observed accuracies were significantly above chance. For each permutation, we randomize the identities of the subjects in both sessions, perform the identification procedures and record the accuracies. A significant level of p-value = 0.05 is used as the threshold for the $10,000$ permutation tests.

\par We then investigated the identification accuracy on the basis of each specific functional network to figure out which brain network contributes more to the individual discriminability. These functional networks are defined in the section of data preprocessing. During this process, a single network or a combination of networks are used to estimate the individual identification. Note that,  if we denote the set of nodes belonging to network $j$ as $V_j = {v_{jk}, k = 1,2,\ldots,K_j}$, where $K_j$ is the total number of nodes in network $j$, only connections within the selected network are included.

\subsection{Fundamental neural activities contribution to individual identification}
\par To check the hypothesis that common neural activities may weaken the individual variability, we compare the performance of identifiability with and without the AE network processing. For this purpose, we exclude the SDL model in this experiment to avoid its influence. As a first pass, we calculate correlations between connectivity matrices of all participants across 9 possible combinations ($s1\rightarrow s2$) under these two scenarios (with and without the AE network). For each correlation matrix, the row and column are symmetric. Thus, diagonal elements are similarity scores from the matched subjects, while off-diagonal elements are the ones from the unmatched participants. By observing the difference between the mean values of diagonal and off-diagonal factors, the individual identifiability can be evaluated. The larger the difference, the stronger the discriminative power.

\par We then estimate the identification accuracy for these two scenarios, respectively. The procedures have already been provided in the section of individual identifiability analysis. If the identification rates generated by the scheme with the AE network are much higher than those without the AE network, we assume that using AE can therefore increase the subject-specific identifiability by reducing the effect of the common neural activities. Finally, we reverse session 1 and session 2 and repeat the above proceedings ($s2\rightarrow s1$).

\par Afterwards, we investigate the regions in the FC, to which the signals removed by the AE network belong. To filter out the influence induced by activities in the task runs, we restrict the analysis to resting-state fMRI (R1 and R2). Group differences between the analyses with and without AE are considered here. First, we calculate the correlation matrix for each subject. Next, we transform correlation matrices into the edge-weight vectors ($e_i$), and concatenate them into the data $Y$ as mentioned in the previous section. Finally, a two-sample t-test is applied to the data $Y$ with a significant level of $q = 0.01$ to examine the group differences.

\subsection{Edgewise contribution to identification}
\par To investigate which connections of the FC contribute more to subject identification, we estimated the modified differential power (DP) provided by Liu et al.\cite{liu2018chronnectome}. In this part, we also restrict the study to R1 and R2 sessions. The modified differential power is defined as follows:

\begin{equation}
\begin{aligned}
    &DP(i,j) = 1-\sum_lP_l(i,j),\\
    &P_l(i,j)=\frac{|\phi_{lk}(i,j)>\phi_{ll}(i,j)|+|\phi_{kl}(i,j)>\phi_{ll}(i,j)|}{2(N-1),}
    \label{equation5}
\end{aligned}
\end{equation}
where $P_l(i,j)$ is an empirical probability to quantify the differential power of an edge for the purpose of subject identification; $l$ and $k$ ($l\neq k$) represent the labels of two different participants; $i$ and $j$ ($i\neq j$) denotes two different nodes within the functional connectivity; $n$ is the total number of participants in the analysis ($n=862$). $|\phi_{lk}(i,j)>\phi_{ll}(i,j)|$ indicates the probability that $|\phi_{lk}|$ between two different subjects is higher than $|\phi_{ll}|$ of the same participant. Given two sets of connectivity matrices $[X_l^{R1}(i,j)], [X_k^{R2}(i,j)]$ obtained from the R1 and R2 sessions after z-score normalization, the corresponding edge-wise product vector $\phi_{lk}(i,j)$ can be calculated as follows:

\begin{equation}
\begin{aligned}
    \phi_{lk}(i,j) = X_l^{R1}(i,j)* X_k^{R2}(i,j), l,k = 1,2,\ldots,N
    \label{equation6}
\end{aligned}
\end{equation}
$|\phi_{ll}|$ can also be obtained in the same way. DP reflects each edge's ability to distinguish an individual subject. For a given functional connectivity, a higher DP value means a greater contribution to individual identification. Furthermore, to investigate the network-dependent contribution to subject-specific identification, we also count the number of the highest DP values (top 1\%) within or between functional networks. In this manner, we examine whether specific brain networks play a significant role in discriminating individuals.

\subsection{Prediction analysis for individual cognitive behavior}
\par To determine whether our refined FC applying the AE network could benefit individual cognitive prediction, we depicted it from two aspects: regression and classification analysis for continuous and discrete targets, respectively. Here, we select items of high-level cognition from the HCP protocol, including fluid intelligence (Penn Progressive Matrices, HCP: PMAT24\_A\_CR, Mean$\pm$SD: 17.04$\pm$4.71, Range: 4-24), cognitive flexibility/executive function (Dimensional Change Card Sort, HCP: CardSort\_AgeAdj, Mean$\pm$SD: 102.54$\pm$9.89, Range: 57.79-122.65), inhibition/executive function (Flanker task, HCP: Flanker\_AgeAdj, Mean$\pm$SD: 102.05$\pm$9.94, Range: 72.81-123.56) and language/vocabulary comprehension (Picture Vocabulary Test, HCP: PicVocab\_AgeAdj, Mean$\pm$SD: 109.44$\pm$15.07, Range: 68.68-153.09). More details can be found at the HCP website (\url{https://db.humanconnectome.org/}). Then, the refined FCs from the R1 session are applied as the features to define these high-cognition scores.


\subsubsection{Regression analysis}
\par To determine whether the refined FC profiles can better predict the individual high-cognitive behavior relative to those without employing the AE network, we use leave-one-subject-out cross-validation (LOOCV) strategy to estimate the prediction accuracy \cite{cai2019refined}. For instance, to assess the ability of refined FCs to describe fluid intelligence, in each LOOCV fold, one participant is assigned as the test sample, and the remaining $n-1$ subjects are considered as the training samples. First, we concatenate all the connections within the FC profiles (i.e., 35778 connections) to generate a feature vector for each subject. Second, we investigate a feature selection step, which calculates the correlation between each connection of FCs (Pearson correlation between two ROIs) and fluid intelligence scores on the training set. If the correlation is significant (p-value $<$0.001), the corresponding feature is retained. Third, a predictive model is built using a simple regression model to fit the selected features to the fluid intelligence coefficients in the training set. Finally, we adopt the model on the unseen test data to generate the behavioral score. During this procedure, each participant is used as the test sample once. After all the LOOCV folds are completed, we assess the predictive power through the correlation values between the predictive and observed fluid intelligence scores.

\par At the end, we perform the permutation test (10,000 times) to test the statistical significance of the observed behavioral scores. For each permutation, the observed behavioral scores of the subjects are randomly shuffled before the regression analysis. In this way, we can examine whether the prediction performance is obtained by chance. 

\subsubsection{Classification analysis}
\par To further evaluate predictive power of refined FCs, we performed classification analyses for two subsets based on these high-cognitive scores.
More precisely, we first extract participants who are within either upper or lower $\delta$-th percentile of the distribution of behavioral scores. That is, we retain subjects who have the highest or lowest $\delta\%$ high-cognitive scores. Cases of $\delta \in \{10, 20, 30\}$ are considered here. Next, the feature selection step discussed in the regression analysis section is applied (the significant level p-value$=$0.001). A support vector machine (SVM) with a Gaussian kernel is used to do the classification.

\par After obtaining two subsets under the different values of $\delta$, a relatively low number of subjects are left for each case (172 subjects for $\delta = 10$, 344 subjects for $\delta = 20$ and 516 subjects for $\delta = 30$). Hence, we repeat the experiment $100$ times. For each run, we divide participants into a training set (75\%) and a testing set (25\%). A SVM function built in Matlab with a Gaussian kernel is employed, and a grid search is applied to optimize parameters within the SVM model (e.g., the radius of the Gaussian kernel, the weight of the soft margin cost function). To validate that reducing the common neural activities can benefit individual discrimination, the results generated by the framework without the AE network are also provided.

\section{Results}
\subsection{Refined FC based individual identification}
\par
\begin{figure*}[htbp]
    \centering
    \includegraphics[width=0.9\textwidth,height=0.4\textwidth]{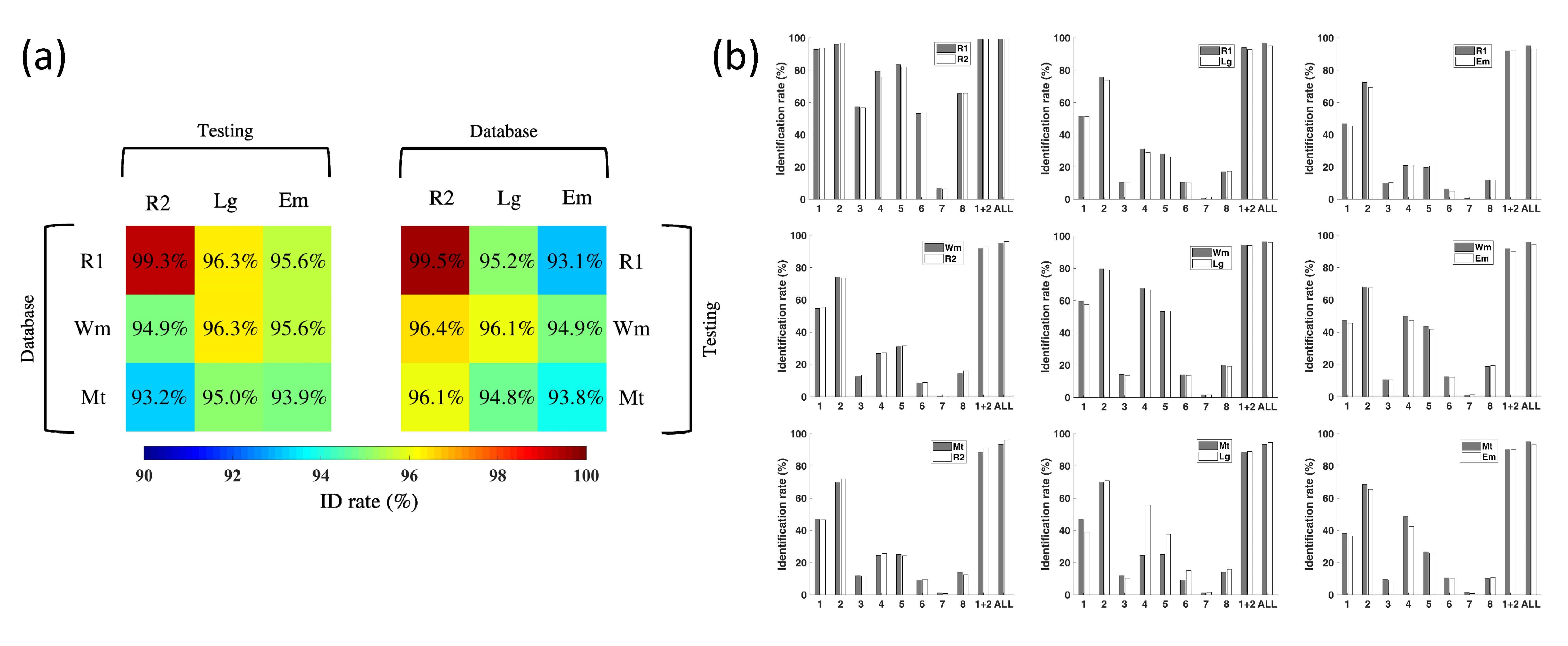}
    \caption{Identification accuracy across session pairs and networks. (a) Identification rates from the whole brain are highlighted in color-coded matrices to compare the accuracies across rest-rest, rest-task and task-task sessions, respectively. R1, Rest1; Wm, Working Memory task; Mt, motor task; R2, Rest2; Lg, language task; Em, emotion task. (b) Identification results based on all 9 sessions in the database and target combinations. Each row shares the same database session and each column shares the same target session. The color of the bar (grey or white) indicates which fMRI modality was used as the database, and the other one was served as the target. Graphs display the identification rate based on each network as well as the combination network 1 and 2 and the whole brain(all).}
    \label{figure3}
\end{figure*}

 As a first pass, we evaluated the identification accuracy applying the whole brain connectivity matrices (268 nodes, without prior network definitions) to validate that the refined FCs can highlight the subject-specific variability. Identification rates are described in Fig.\ref{figure3}(a). Even with a large number of participants (862 subjects), refined FCs worked well for individual identification. The success rates were $99.3\%$ and $99.5\%$ based on a database-target rest1-rest2 and the reverse rest2-rest1, respectively. Meanwhile, the identification rates ranged from $93.1\%$ to $96.4\%$ for rest-task pairs and $93.8\%$ to $96.3\%$ for task-task combinations. In relative to raw FCs ($900$ subjects, $89\%$ accuracy between the two rest sessions, $65\pm14\%$ for other pairs in Finn et al.'s work \cite{finn2017can}), refined FC profiles largely improved the performance of individual discrimination. Given that identification trials were not independent of each other, we performed 10,000 nonparametric permutation tests (two-sided) to assess the significance level of these results. Across 10,000 iterations, the p-value for each pair of sessions is below 0.0001. It indicates that the success rates of identification are significantly above chance.

\par Next, we examined the identification accuracy based on each functional network to explore which brain network contributes more to the individual variability. These networks are defined by Finn et al. \cite{finn2015functional} and depicted in Fig.\ref{figure1}. The medial frontal network (network 1) and the frontoparietal network (network 2) achieve the highest success rates in individual discrimination, which comprises the higher-order association cortices in the frontal, parietal and temporal lobes. In comparison with the medial frontal network, the frontoparietal network performs much better especially for the rest-task and task-task pairs. Furthermore, we also checked whether the combination of networks 1 and 2 can provide better performance than each individual. As shown in Fig.\ref{figure3}(b), in each scenario, the identification accuracies using the combination of networks 1 and 2 are higher than implementing network 1 or 2 independently, and are pretty close to those generated through applying the whole-brain nodes. For other networks, subcortical-cerebellum network (network 4) and motor network also contribute to subject-specific variability.

\subsection{Fundamental neural activities contribution to identification}
\par
\begin{figure*}[htbp]
    \centering
    \includegraphics[width=1.0\textwidth,height=0.6\textwidth]{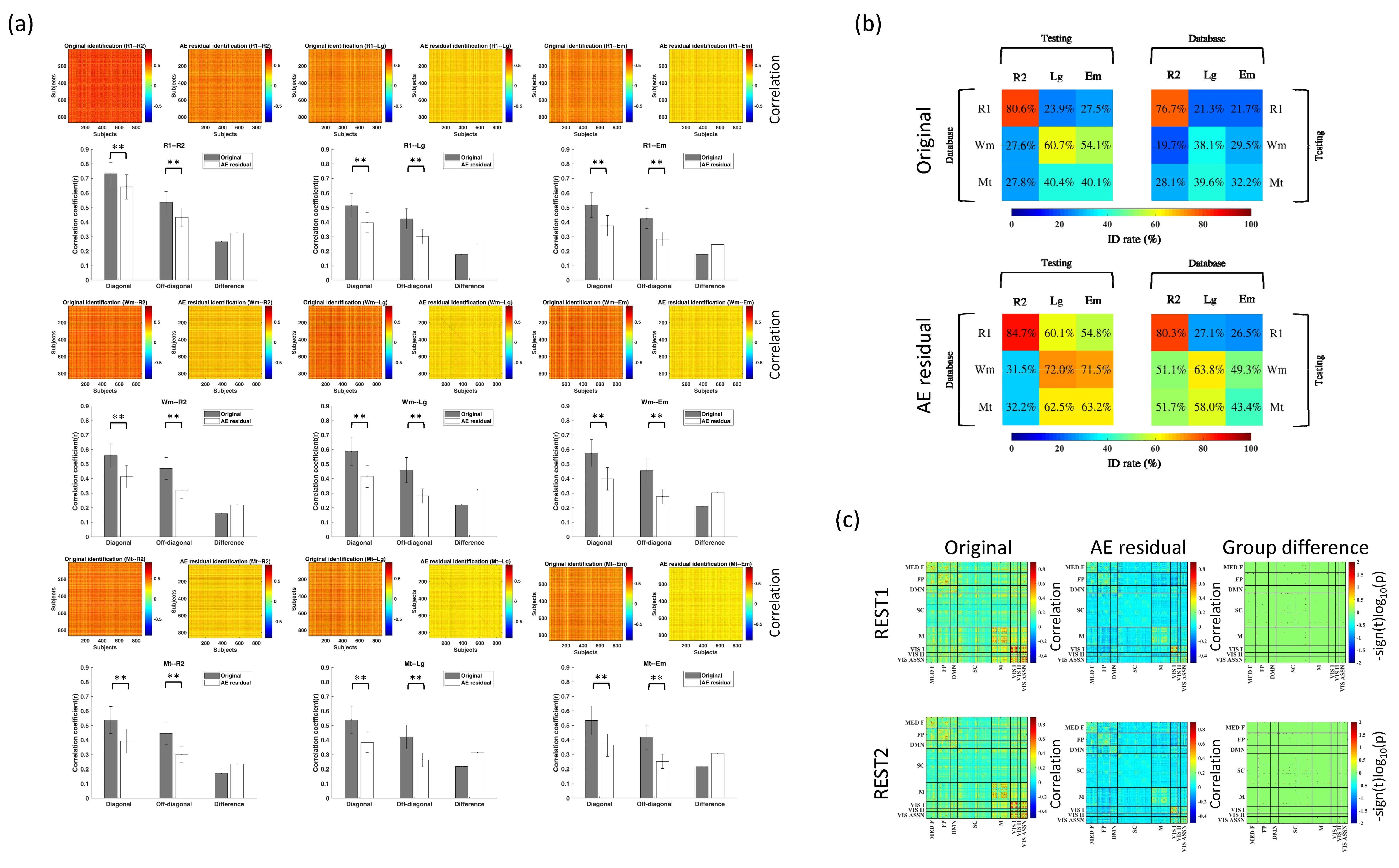}
    \caption{Evaluation of the influence of common neural activities on individual identification. (a) Analysis of identifiability matrices based on all 9 pairs of the combination of a session in the database and the target ($s1\rightarrow s2$). For the top line of each sub-figure, from left-right: identifiability matrix (i.e. Pearson correlation coefficient between functional connectivity across subjects and modalities) of the original data; identifiability matrix of the AE residual data. The row and column subject order of identifiability is symmetric. Hence, diagonal elements are correlation scores from the matched subjects, while off-diagonal coefficients are from the unmatched participants. Mean correlation coefficients for both diagonal (match) and off-diagonal (unmatched) elements are also calculated  (bottom, err bars indicates $\pm$ s.d.). ** means p-value$<$ $10^{-5}$ for two-tailed t-test. Besides, the difference (mean value) between the diagonal and off-diagonal elements for both the original and AE residual data are displayed. (b) Comparison of identification accuracies across all 9 pairs of database session and target session between using original time series (top) and AE residual data (bottom). Note that only the situation with whole-brain nodes (264 nodes) is considered here. (c) Static functional network connectivity for the original data (left) and AE residual signals (middle) estimated by the Pearson correlation. Meanwhile, group differences (right) between them is obtained by applying a two-sample t-test. These variations are visualized by plotting the log of p-value with the sign of t statistics, $-sign(t)log_{10}(p)$. Note that in this part, the results from the rest1 and rest2 are displayed individually. }
    \label{figure4}
\end{figure*}

 To investigate the contribution of common neural activities to individual identifiability, we estimated the correlations between connectivity matrices of all subjects across 9 possible combinations for both time courses with and without the AE network processing. For each correlation matrix, the row and column are symmetric by subject. Thus, diagonal elements are correlation coefficients from the matched subjects, while off-diagonal elements are those from the unmatched participants. The results for all 9 possible pairs are displayed in Fig.\ref{figure4}(a). In comparison with raw cross-subject correlation coefficients, scores generated after the AE network significantly become weak for both diagonal and off-diagonal factors. However, applying the AE network improves the difference between diagonal and off-diagonal elements in the correlation matrix. It indicates that reducing the contribution from common neural activities mentioned above indeed helps individual discrimination.

\par Next, we repeated the identification experiments applying connectivity matrices from the two scenarios discussed above to further validate our hypothesis. By checking the identification results in Fig.\ref{figure4}(b), we observe that identification accuracies for all 9 pairs have increased through reducing the common neural contribution. More specifically, for the pairs of rest-rest, the identification rates were improved around $4\%$. As to rest-task combinations, when using connectivity matrices of resting-state fMRI in the database, the ability of discrimination has significantly enhanced (R1-Lg : $36.2\%$, R1-Em : $27.3\%$, R2-Wm : $31.4\%$, R2-Mt : $23.6\%$). By contrast, the rates applying matrices achieved from task-based fMRI in the database gain around $4\%$. Meanwhile, with the AE network preprocessing, the identification accuracies increase for every condition of task-task combination, ranging from $11.2\%$ (Em-Mt) to $25.7\%$ (Lg-Wm). Thus, weakening the signals from common neural activities can significantly enhance inter-subject differences, and the pairs of rest-rest possess the strongest identification power.

\par To explore which connections were removed by the AE networks, we tested the group average functional connectomes before and after applying the AE networks and examined the difference between them using a two-sample t-test. From Fig.\ref{figure4}(c), we obtain that the strength of links in the connectomes reduces overall. However, the significant difference is largely related to the frontoparietal and subcortical-cerebellum networks.

\subsection{Evaluation edgewise contributions to identification}
\par To  determine which connections contribute more to subject-specific identification, we calculated the modified differential power (DP). The modified DP reflects each edge's ability to distinguish an individual from a pool of participants. A connection with high DP tends to have a similar value within an individual across modalities, but possess a different degree across individuals regardless of modalities.

\par
\begin{figure*}[htbp]
    \centering
    \includegraphics[width=0.9\textwidth,height=0.4\textwidth]{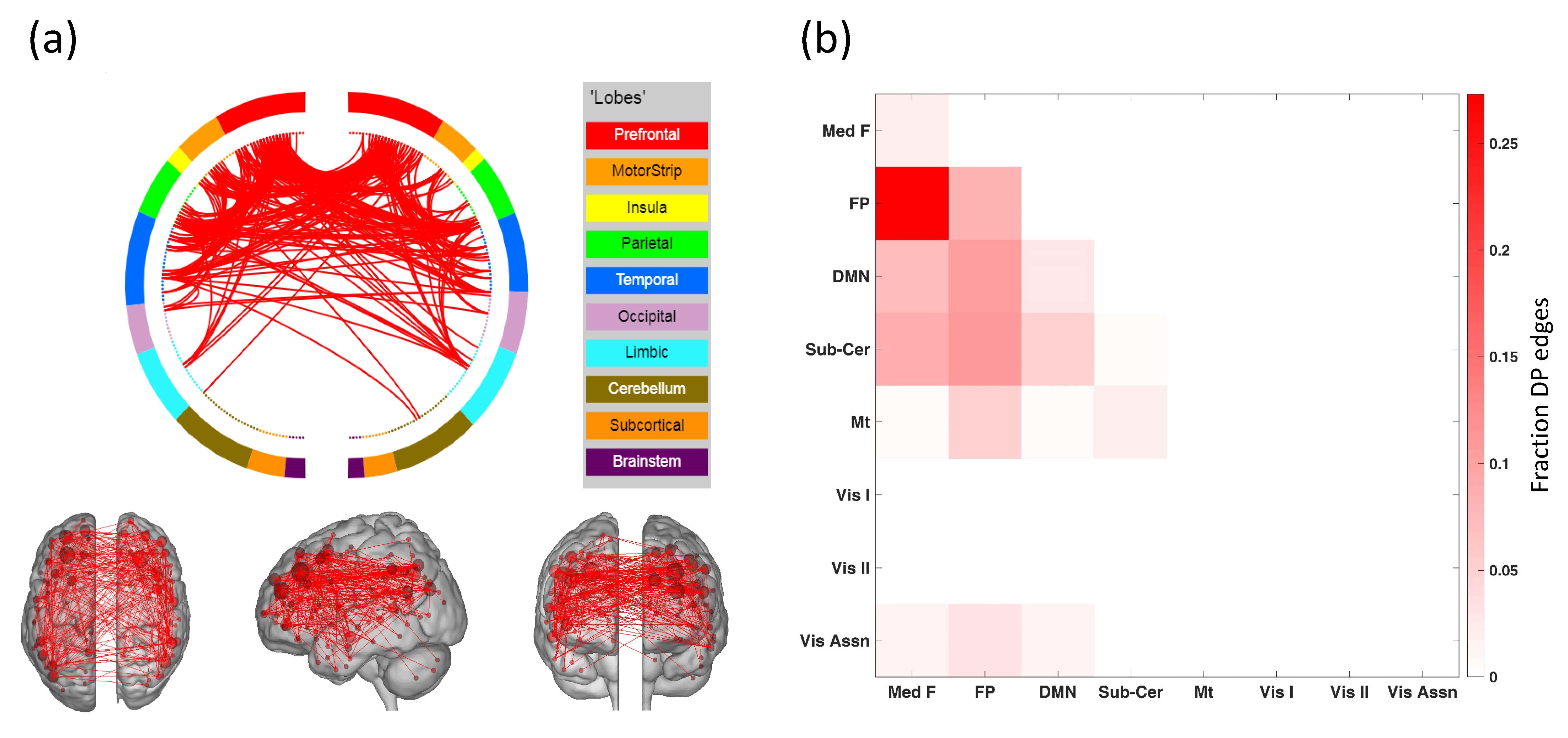}
    \caption{Edgewise contributions to individual identification. (a) Connections that possess the highest DP scores in individual connectivity profiles (top, circle plot). Axial, sagittal and coronal views of these links are also provided (bottom, from left to right). Note that connections with the highest 1\% DP values are shown here.  In the circle plots (top), the 268 nodes (the inner circle) are organized into a lobe scheme (the outer circle) roughly reflecting brain anatomy from anterior (top of the circle) to posterior (bottom of the circle) and split into left and right hemisphere. Lines indicate edges or connections. (b) The percentage of connections within and between each pair of networks (8 functional networks defined in Fig.\ref{figure1}) using the same data as (a). The color depth of the grid in the matrix indicates the fraction of DP edges for each pair of networks.  }
    \label{figure5}
\end{figure*}

 By restricting the analysis to resting-state fMRI (rest1 and rest2), we estimated the modified DP for all edges in the brain. We determined which connections were in the 99.9 percentile across all of the links (Fig.\ref{figure5}). We observe that the majority of edges in the 99.9 percentile of the edges are in the frontal, parietal, and temporal lobes. Meanwhile, most of the nodes with high DP values involve in the frontoparietal and medial frontal networks. Some of them belong to the default mode network. By checking the results in Fig.\ref{equation5}(b), we obtain that for the connections possessing high DP values in the connectivity, $27.3\%$ of them are used to link the medial frontal and frontoparietal networks. In addition, $59\%$ are connections linking these networks to others ($38\%$ of them link the frontoparietal network to other networks). It indicates that connections related to high-order association cortices are the most discriminative of individuals. On the other hand, medial frontal and frontoparietal networks play a significant role in individual identification.

\subsection{Connectivity profiles predict high-level cognitive behaviors}
\subsubsection{Regression analysis}
\par
\begin{figure*}[htbp]
    \centering
    \includegraphics[width=0.9\textwidth,height=1.2\textwidth]{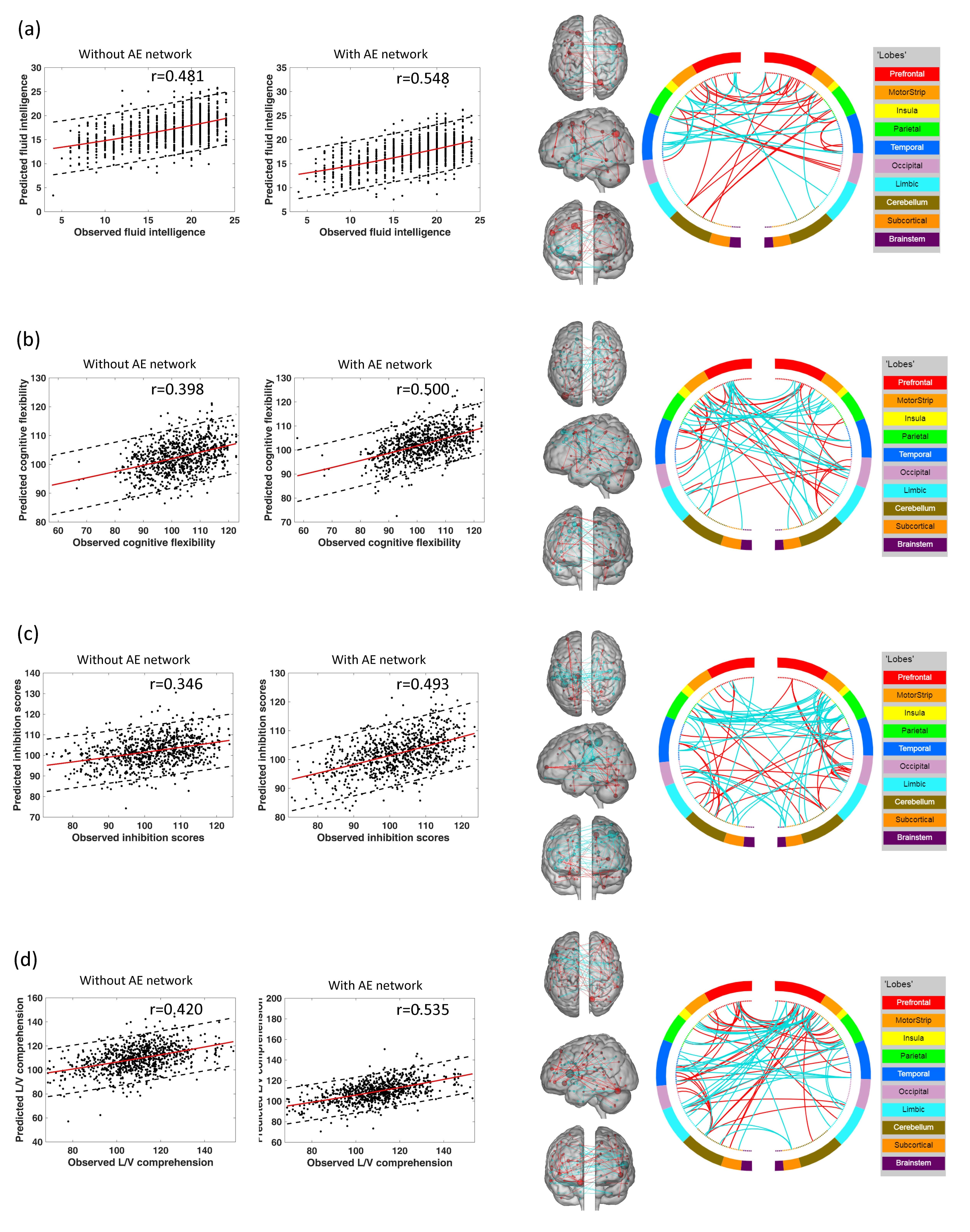}
    \caption{Connectivity profiles predict cognitive behavior. Scatter plots display prediction results from a leave-one-subject-out cross-validation (LOOCV) analysis comparing the predicted and the observed high-level cognitive scores. Both the connectomes with and without applying the AE network processing are considered. Note that under these two scenarios, the sparse dictionary learning (SDL) model is included. In the scatter plot, each dot represents one subject, and the area between dashed lines reflects $95\%$ confidence interval for the best-fit line, which is used to assess the predictive power of the model. R-values are the correlation coefficients between the predicted and observed high-level cognitive scores. Furthermore, for each cognitive scores, edges retained from the feature selection step (p-value$<$0.001) are also depicted (with the AE network process). Axial, sagittal, and coronal views of these connections in the brain are provided. In the circle plots, the 268 nodes (the inner circle) are organized into a lobe scheme (the outer circle), roughly reflecting brain anatomy from anterior (top of the circle) to posterior (bottom of the circle), and divided into left and right hemisphere. Red and blue lines mean positive and negative connections, respectively. (a) fluid intelligence; (b) cognitive flexibility/executive function; (c) inhibition/executive function; (d) language/vocabulary comprehension.}
    \label{figure6}
\end{figure*}

 To test whether refined FC profiles benefit the behavior prediction, we explored the prediction abilities across different high-level cognitive scores of connectivity profiles under the scenarios with and without the AE network processing. Note that for these two conditions, the SDL model was included during the analysis. As demonstrated in the scatter plots in Fig.\ref{figure6}, the predicted scores by the connectivity profiles with AE network have higher correlation coefficients with the observed scores relative to those without the AE network. Besides, the range of predicted scores by the refined FCs is much narrower across all the cognitive scores (especially for the language/vocabulary comprehension). To validate the prediction power of our proposed framework applying the AE network, we performed 100 nonparametric permutations for each score. The results illustrate that the prediction of each high-level cognitive behavior (correlation between observed and predicted scores) is above chance (fluid intelligence: p-value$<$0.01; cognitive/executive flexibility: p-value$<$0.08; inhibition/executive function: p-value$<$0.06; language comprehension: p-value$<$0.01). It means that reducing the common neural activities also helps predict cognitive behavior.

\par Furthermore, by observing the selected features from refined FC profiles, we find that different brain regions exhibit distinct contributions to various high-level cognitive parameters. More specifically, frontal and parietal lobes contribute most to fluid intelligence prediction according to this study. In detail, nodes located in the frontal and parietal regions provide positive connections highly related to fluid intelligence. Moreover, some of the positive links are relevant to the right cerebellum, while a large portion of negative links connects with the insula region. When using the refined FC profiles to predict the performance of cognitive flexibility, in comparison with positive edges, negative connections play a more critical role. Most of these negative edges are linked with frontal and parietal lobes. Notably, the region of the motor strip is involved in both the positive and negative connections regarding cognitive flexibility. Similarly, negative connections predominantly contribute to inhibition function as well. Among these negative links, the majority of them interact with the parietal lobe. Also, the negative edges from the motor strip and subcortical regions have equivalent influence on the prediction of inhibition function. In the prediction of language comprehension, through both the positive and negative edges in the connectivity, the frontal and temporal lobes closely connect with language comprehension. As a consequence, the data-driven framework applying the refined FC profiles can adequately enable us to find out which brain regions closely interact with high-level cognitive behavior.

\subsubsection{Classification analysis}
\par To analyze the prediction power of refined FCs further, we investigated whether we were able to discriminate two subsets based on high-level cognitive behavior. In the experiments, the results with and without the AE network processing were compared. In general, regardless of the percentile values $\delta$, applying refined connectomes leads to satisfactory classification rates for all the cognitive scores ($>75\%$). However, reducing the common neural activities with the AE network has a different influence on the classification of various cognitive parameters. More specifically, relative to only implementing the SDL model, the proposed refined FCs have better performance in predicting fluid intelligence. While these two frameworks offer similar classification rates with $\delta=10$ (mean values, with the AE network: $78.3\%$, without the AE network: $81.1\%$), refined FC profiles possess more stable classification accuracies as the difference of fluid intelligence between the two subsets decreases. Interestingly, functional connectomes generated by implementing the AE network showed an improvement in classifying cognitive flexibility related subsets. Regarding those without the AE network, the accuracy increases from $77.5\%$ to $81.6\%$ for $\delta=10$, $74.7\%$ to $78.3\%$ for $\delta=20$ and $70.7\%$ to $74.3\%$ for $\delta=30$. As to the cognitive parameters of inhibition function and language comprehension, while the slight differences exist, whether adopting the AE network or not will not affect the performance of classification.

\begin{figure*}[htbp]
    \centering
    \includegraphics[width=0.9\textwidth,height=0.4\textwidth]{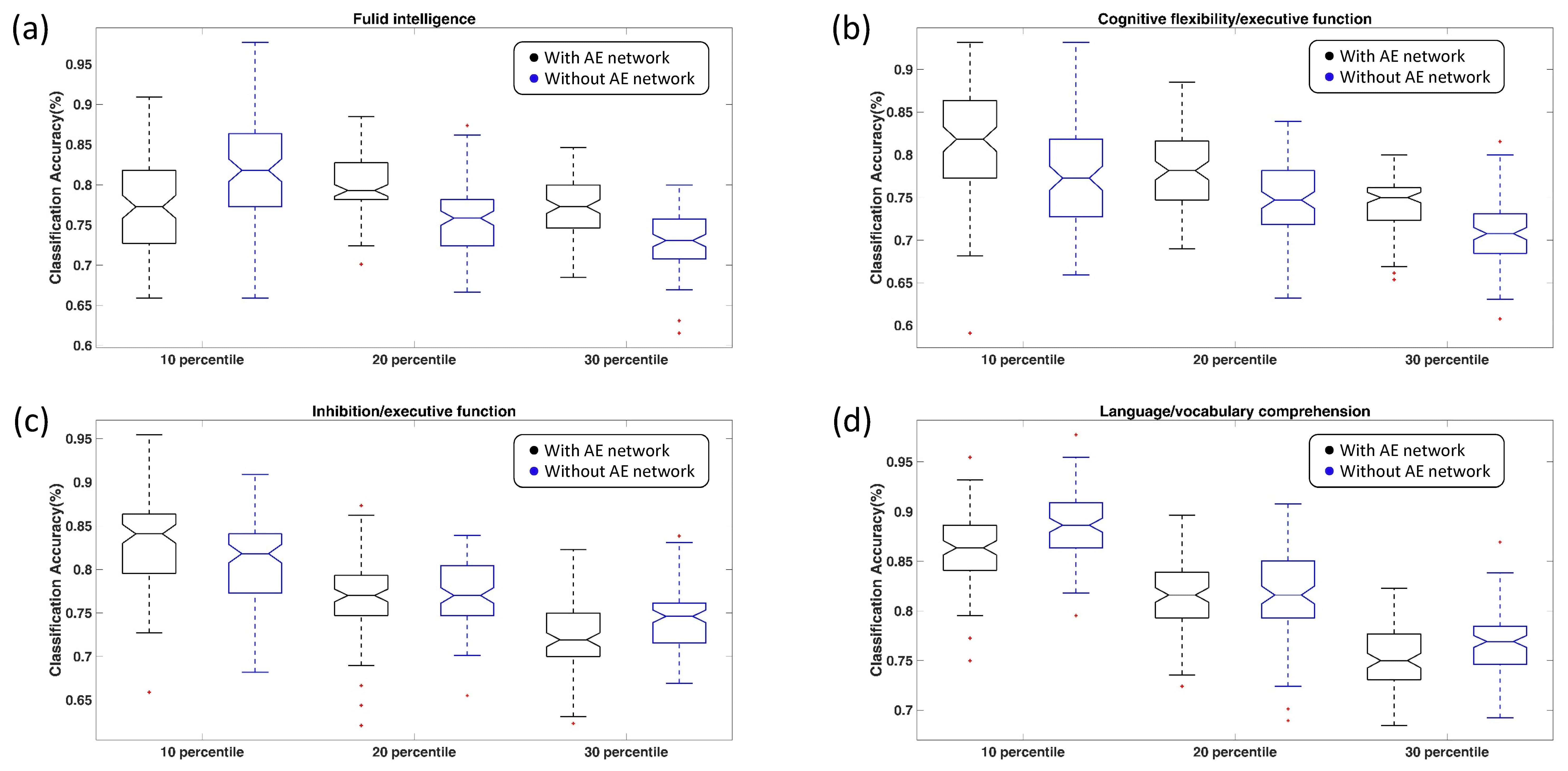}
    \caption{Classification results between low and high cognitive groups based on high-level cognitive behavior. Through the feature selection step, significant edges are retained (p-value$=$0.001). Then, three percentile values are considered ($\delta=10, \delta=20\ and\ \delta=30$). Black boxes provide the classification accuracies applying FC profiles with the AE network processing, while blue boxes represent the rates using those without the AE network. Note that in the analysis, the SDL model is involved in these two different scenarios.}
    \label{figure7}
\end{figure*}

\section{Discussion}
\par Recently, the study and use of dynamic functional network analysis has drawn more attention. Some basic configurations in brain connectivity appear across all time-varying states, and may not contribute to subject-specific discrimination. In our previous work, we pointed out that refining the measures of individual connectomes can help improve the fingerprinting power of individuality \cite{cai2019refined}. In this study, we applied the autoencoder network to extract the signals of common neural activities, which may hinder the identification of individual difference. Therefore, we removed them from the raw time courses when constructing connectivity network between ROIs. We then applied dictionary learning on all the data. Based on the pipeline, we generated the new refined FC connectomes \cite{cai2019refined}. We showed that the refined FC profiles can successfully distinguish one individual from a pool of the population with a high identification accuracy. Moreover, the refined connectomes can significantly predict high-level cognitive behavior, including fluid intelligence, cognitive flexibility, inhibition function, and language comprehension. Notably, reducing the signals of common neural activities benefited individual identification and cognition prediction, where frontal, parietal and temporal lobes contributed significantly. Collectively, our findings supported our assumption that common neural activities may impede the individual identifiability and its removal can help enhance the uniqueness of each individual.

\par When testing the identification ability of refined FCs, we found that regardless of database-target combinations, the connectomes applying the proposed approach successfully distinguished each individual from all the participants. The accuracy was from $99.3\%$ to $99.5\%$ for the rest-rest pairs. As to the rest-task and task-task combinations, the success rates ranged from $93.1\%$ to $96.4\%$ and from $93.8\%$ to $96.3\%$, respectively. The strong individual identification power of refined FCs from our results suggest that connectomes vary across participants and are unique for each subject. This is in agreement with previous findings that the connectomes could be used as the fingerprint to identify an individual \cite{finn2015functional, cai2019refined}. Compared to results obtained by raw FCs, reducing the effect of common neural activities and group factors can significantly improve the success rates of identification \cite{finn2017can}. For the pairs of rest-rest fMRI, we gained $10\%$ accuracy improvement. Also, the success rates increased by around $30\%$ for other combinations. The findings validate our assumption that some patterns caused by common neural activities may impede the individual variability, and thus reducing their impact helped capture unique characteristics of the brain.

\par The contributions of functional networks to individual identification was also examined. Although discrimination based on the whole connectivity matrix performed best, the medial frontal and frontoparietal networks achieved high accuracy. The combination of these two networks provided better performance than only one of them alone. These networks are composed of higher-order association cortices (fontal, parietal and temporal lobes), which have been proven to show the highest inter-subject variance \cite{finn2015functional, cole2013multi}. Relative to the medial frontal network, frontoparietal contributed more to the identification, especially for rest-task and task-task pairs. It is consistent with the function of the frontoparietal system, which is particularly active in tasks requiring a high degree of cognitive control. Even for the rest-rest combination, the frontoparietal network worked very well for identification. Thus, we believe that the frontoparietal system plays a significant role in the brain's uniqueness regardless of whether the mind is at rest or not. Also, we detected that the subcortical-cerebellum and motor network positively correlated with individual differentiation. It matches with the conclusion that there was a gradual increase of variability in primary regions of the visual and sensorimotor systems specific to subcortical and cerebellum structures as the brain developed \cite{li2017linking}.

\par To further estimate the contribution of common neural activities to individual variability, we compared the correlation matrices with and without AE network, and repeated the identification experiments for these two cases. While applying the AE network weakened both diagonal and off-diagonal factors of correlation matrices, reducing the signals from common connectivity patterns increased the difference between diagonal and off-diagonal elements. Besides, without using the SDL model, reducing the common neural contribution increased the identification rates across all 9 combinations. All these findings indicate that weakening the contribution of common neural activities helps strengthen the subject-specific variability and enhances the uniqueness of the human brain. Intriguingly, we observed that the time courses identified by the AE networks are mainly related to the frontoparietal and subcortical-cerebellum systems, which had a significant influence on the identification of individuals. In light of this, the interaction between these networks with individual predictions will be conducted in future work.

\par By examining the modified differential power of each edge in the connectivity map, we determined that most of the connections with high DP values were related to the higher-order association (frontal, parietal, and temporal) lobes. Also, $27.3\%$ of high DP edges were connected with the medial frontal and frontoparietal systems, and $57.3\%$ of them linked these networks to others. These findings further validate the function of the medial and frontoparietal networks in individual identification.

\par When exploring the prediction abilities across different cognitive parameters with and without AE processing, we observed that the predicted scores by applying AE processing possessed higher correlation coefficients with the observed scores relative to those without AE processing. This suggests that the reduction of  common functional patterns can improve the power of cognitive behavior prediction. Next, we analyzed the selected features that contributed more to the cognition prediction and found that different brain regions had various effects on each cognitive measurement. We demonstrated large contributions of the frontal and parietal lobes to individual fluid intelligence and execution function (e.g., flexibility and inhibition function). These findings are consistent with previous studies \cite{cole2013multi, finn2015functional, liu2018chronnectome}. Specifically, we detected that insula was closely associated with fluid intelligence. This agrees with the statement that fluid intelligence has been correlated with a distributed network comprising regions of frontal, insula and parietal cortex \cite{tschentscher2017fluid}. Furthermore, we found that the motor strip region was highly related to executive function. This point is supported by the argument that motor and cognitive processes are functionally related and most likely share a similar evolutionary history. It is well established that  multiple brain regions integrate both motor and cognitive functions \cite{leisman2016thinking}. For language comprehension, we demonstrated that the frontal and temporal lobes closely interacted with it. Broca' area (in the frontal lobe) and Wernicke's area (in the temporal lobe) are cortical areas that respond to human language. In sum, connectomes which are refined by our approach can improve our ability to characterize the relationship between brain regions and cognitive behavior and help enhance our understanding of the human brain.

\par To further analyze the predictive power of refined FCs, we investigated the classification analysis and obtained satisfying classification rates regardless of the percentile values $\delta$. By using the AE network, more stable rates were obtained for fluid intelligence and cognitive flexibility with gained accuracy across all the conditions. However, no improvement was made by removing signals from neural activities for the language comprehension. Hence, we consider the basic configuration in the connectivity map has a different contribution to various cognitive behavior.

\par Several issues need further consideration. First, in this work, we still implemented the static functional network connectivity. However, recent research has noted that dynamic functional connectivity could provide complementary individual information \cite{liu2018chronnectome}. A combination of dynamic and static connectivity is a promising direction for analyzing individual variability. Second, several confounding factors might affect the performance of our proposed approach, such as parcellation schemes, and head motion. Follow-up studies are needed to further explore these factors. Third, we focus on group common and individualized aspects of the connectome. More work is needed to more fully understand these aspects of brain function.

\section{Conclusion}
\par In this work, we assumed that the common neural activities might weaken the difference in brain connectivity across participants. By proposing our framework including the autoencoder network, we reduced these common patterns of connectivity to enhance the uniqueness of each individual. We observed that refined FC profiles estimated by our proposed pipeline can identify each individual with high accuracy (up to $99.5\%$ for the rest-rest pair). Meanwhile, connectomes refined by our approach can also be used to predict high level cognitive behavior (e.g., fluid intelligence). Hence, reducing the signals of common neural activities indeed improved both individual identification and prediction of cognitive function, where frontal, parietal, and temporal lobes contributed significantly. In summary, the findings in this study validate our hypothesis, and our proposed approach provides a promising way to study individualized brain networks.

\section*{Acknowledgment}
\par Data were provided in part by the Human Connectome Project, WU-Minn Consortium (principal investigators, D. Van Essen and K. Ugurbil; 1U54MH091657) funded by the 16 US National Institutes of Health (NIH) institutes and centers that support the NIH Blueprint for Neuroscience Research; and by the McDonnell Center for Systems Neuroscience at Washington University. The authors would like to thank the partial support by NIH (R01\ GM109068, R01\ EB020407,
R01\ MH104680, R01\ MH107354, R01 MH103220) and NSF (\#1539067).

\ifCLASSOPTIONcaptionsoff
  \newpage
\fi

\bibliographystyle{IEEEtran}
\bibliography{reference}

\end{document}